# Decoding the Mechanisms of Reversibility Loss in Rechargeable Zinc-Air Batteries

Zhibin Yi[1,2], Liangyu Li[2], Cheuk Kai Chan[3], Yaxin Tang[1], Zhouguang Lu[1], Chunyi Zhi[4], Qing Chen[2,3]*, and Guangfu Luo[1,5]*

[1]Department of Materials Science and Engineering, Southern University of Science and Technology, Shenzhen 518055, P. R. China

[2]Department of Mechanical and Aerospace Engineering, The Hong Kong University of Science and Technology, Clear Water Bay, Hong Kong, P. R. China

[3]Department of Chemistry, The Hong Kong University of Science and Technology, Clear Water Bay, Hong Kong, P. R. China

[4]Department of Materials Science and Engineering, City University of Hong Kong, Kowloon, Hong Kong, China.

[5]Guangdong Provincial Key Laboratory of Computational Science and Material Design, Southern University of Science and Technology, Shenzhen 518055, China

*E-mail: chenqing@ust.hk, luogf@sustech.edu.cn

## Abstract

Attaining high reversibility of electrodes and electrolyte is essential for the longevity of secondary batteries. Rechargeable zinc-air batteries (RZABs), however, encounter drastic irreversible changes in the zinc anodes and air cathodes during cycling. To uncover the mechanisms of reversibility loss in RZABs, we investigate the evolution of zinc anode, alkaline electrolyte, and air electrode through experiments and first-principles calculations. Morphology diagrams of zinc anodes under versatile operating conditions reveal that the nano-sized mossy zinc dominates the later cycling stage. Such anodic change is induced by the increased zincate concentration due to hydrogen evolution, which is catalyzed by the mossy structure and results in oxide passivation on electrodes, and eventually leads to low true Coulombic efficiencies and short lifespans of batteries. Inspired by these findings, we finally present a novel overcharge-cycling protocol to compensate the Coulombic efficiency loss caused by hydrogen evolution and significantly extend the battery life.

**Introduction**

Zinc-air batteries have a longstanding history as primary batteries, powering small electronic devices such as hearing aids, due to their high energy densities, low cost, and excellent safety in aqueous electrolytes. These inherent advantages have also sparked significant interest in the development of rechargeable zinc-air batteries (RZABs), with potential applications ranging from safe and large-scale energy storage to extreme working conditions[1-4]. However, the widespread commercialization of RZABs still faces a significant challenge of their short cycle lives[5]. Although the implementation of tanks of flow electrolytes can mitigate this issue, it comes at the expense of energy densities[6]. Increasing the cycle lives remains a critical objective in the pursuit of practical RZABs.

The short cycle lives of RZABs originate from challenges in both the air cathode and Zn anode. In the cathode, inadequate catalytic activities of catalysts push the charging voltage towards extreme values that degrade both the catalysts and current collectors during oxygen evolution[7-11]. Meanwhile, the Zn anodes present well-known irreversible changes during cycling, including shape change, dendrite formation, oxide passivation, and gas evolution[12]. These issues result in capacity and efficiency losses, ultimately leading to premature failures[13]. Previous research on Zn deposition in ZnO-saturated alkaline solutions demonstrated that the Zn morphologies generally transition from mossy to layered and then to dendritic deposits with increasing current density[14-15]. Notably, the mossy deposit, which is rarely seen in other metal depositions, tends to grew preferentially along the [0001] direction[16]. Both the mossy and dendritic structures can easily detach from zinc anode and potentially pierce the separator to cause capacity loss and poor cycling stability[17]. Furthermore, ZnO has been found to nucleate on the Zn anode and air electrode, obstructing further electrochemical reactions[18]. Hydrogen bubbles induced by the hydrogen evolution reaction (HER) on the anode can further reduce the efficiency and increase the overpotential of batteries[19]. To untangle the aforementioned phenomena and enhance the lifespan of RZABs, it is imperative to obtain a comprehensive understanding of the underlying mechanisms.

We devote this work to uncovering the mechanisms behind the entangled phenomena related to the loss of reversibility in RZABs under typical operating conditions. Through a combination of experimental investigations and first-principles calculations, we discover the dynamics of electrode morphologies in relation to current densities, initial zincate concentrations, and cycling states, and reveal the intimate relationships between morphology evolution, the electrolyte compositional changes, and HER. With such understanding, we finally present an innovative overcharge-cycling protocol that significantly prolongs the lifespan of RZABs.

**Results and Discussion**

**I. Dynamics of Zn Morphology**

We use a cell comprising a stack of acrylic plates, the most common type in RZAB research (**Figure S1a**). Polished Zn plates (3 cm$^2$ in the electrode area) were used as anode, a composite of Pt/C and IrO$_2$ was employed as the catalyst in the cathode, and the electrolyte was 6 M KOH with different initial zincate concentrations. The cells were first discharged and then charged with a fixed capacity of 5 mAh/cm$^2$ unless stated otherwise. The constant current density $i$ ranges from 1 to 100 mA/cm$^2$ and the initial zincate concentrations $C_0$ from 0.1 to 0.5 mol/L, conditions covering the typical ones for RZABs in literature (**Table S1**). Representative cycling curves are shown in **Figure S1b-d**, all of which display stable voltages for over 50 cycles, typical of RZABs in literatures with similar builds. At different stages of the cycling tests, we took the anode out for characterization to piece together the morphological evolution. See more details on the experiments in the Supporting Information.

Characterizations of the anodes immediately after the 1$^{st}$ discharging-charging cycle via scanning electron microscopy (SEM) reveal three distinct morphologies: mossy, compact, and dendritic deposits (**Figure 1a**), whose appearances depend on both $i$ and $C_0$. As shown in **Figure 1b**, the mossy deposits, featuring slim and curly Zn filaments of ~90 nm in diameter (**Figure S2**), are formed under low $i/C_0$ ratios. By contrast, high $i/C_0$ ratios lead to the dendritic deposits. The intermediate $i/C_0$ ratios form the compact deposits with micrometer-sized hexagonal crystals. These are all distinct from the initial morphology, shown in **Figure S3**. **Figure 1c** translates **Figure 1b** into a morphology diagram to better visualize the dependence of morphology on $i$ and $C_0$.

The morphology diagram evolves with the number of cycles. After the 5$^{th}$ discharging-charging cycle, the mossy morphology claims more territory in the diagram, and the dendritic morphology takes over the condition of $i = 80$ mA/cm$^2$ and $C_0 = 0.2$ M (**Figure 1d**, with the corresponding SEM images in **Figure S5a**). After the 50$^{th}$ cycle, the region of the mossy morphology further expands, whereas that of the compact morphology continues to shrink (**Figure 1e** and **Figure S5b**).We need to stress that near the mossy-compact and dendrite-compact boundaries in **Figure 1d** and **1e**, the two neighboring morphologies typically coexist in the deposits. In these regions, the deposits are classified by the dominant morphology across regions ~100 μm wide. Additionally, the compact deposits could be subcategorized into ridges and boulders (as shown in **Figure S8c2**) according to Bockris et al[20]. We do not differentiate them here because (i) they nearly always coexist, and the ridges often develop into boulders; and (ii) their differences in porosity and roughness are much smaller, when compared with the mossy and the dendritic deposits.

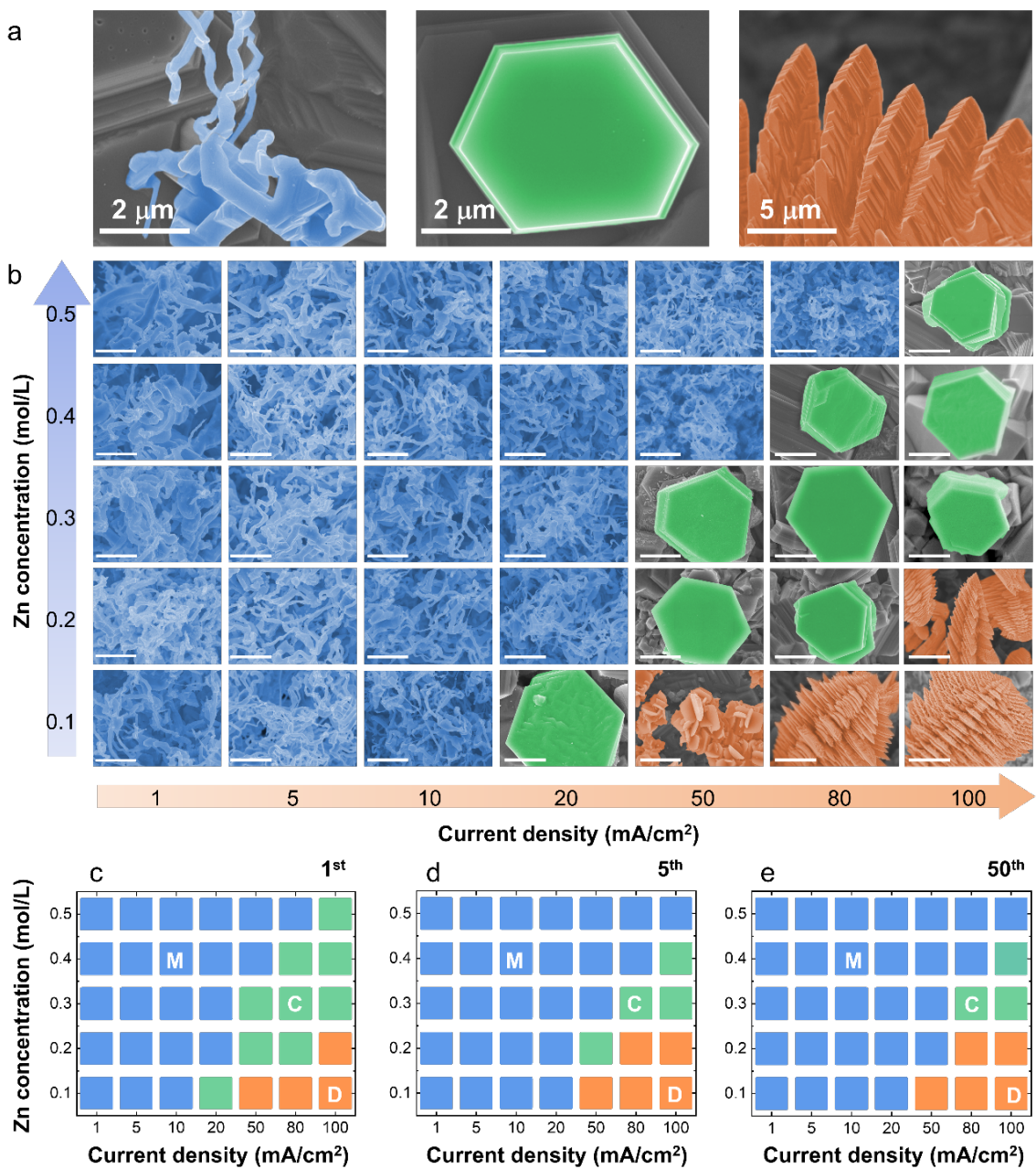

**Figure 1**. Morphology evolution of the Zn anodes. (a) Morphologies of the mossy, compact, and dendritic Zn deposits. (b) Typical SEM images of the Zn anodes after the 1st discharging-charging cycle under different current densities and initial zincate concentrations. The scale bars are 2 μm. (c – e) Morphology diagrams of Zn deposits after the 1st, the 5th, and the 50th cycle, respectively, based on corresponding SEM images (Figure S4-5). Conditions dominated by the mossy, the compact, and the dendritic morphology are colored in blue, green, and orange, respectively. This color code is used consistently throughout the work.

Transmission electron microscopy (TEM) and X-ray diffraction analysis (XRD) reveal the crystal orientations of the three types of deposits, consistent with their shapes. As shown in the high-resolution TEM images, the mossy structure surface exhibits a lattice spacing of 2.6 Å (**Figure 2a**), consistent with that between the (0001) planes. The compact deposit (**Figure 2b**) and the dendritic branch (**Figure 2c**) show a ($10\bar{1}0$) spacing of 2.3 Å, consistent with their hexagonal facets (**Figure S6b** and **S6c**). XRD (**Figure 2d**) further quantifies the fraction of each surface orientation in the deposits. While the polished Zn started with a higher peak intensity of (0001), cycling at 20 mA/cm$^2$ raises the ($10\bar{1}0$) peak substantially due to the mossy deposition. At 50 mA/cm$^2$, the (0001) peak becomes more prominent initially given the compact deposit, but its ratio over the ($10\bar{1}0$) peak gradually decreases given the evolution of mossy deposit as seen in **Figure 1b** and **S5**. As for 100 mA/cm$^2$, the two peaks stay roughly at similar intensities, underlaid by the diffusion-controlled kinetics that does not differentiate the energetics of the facets as much.

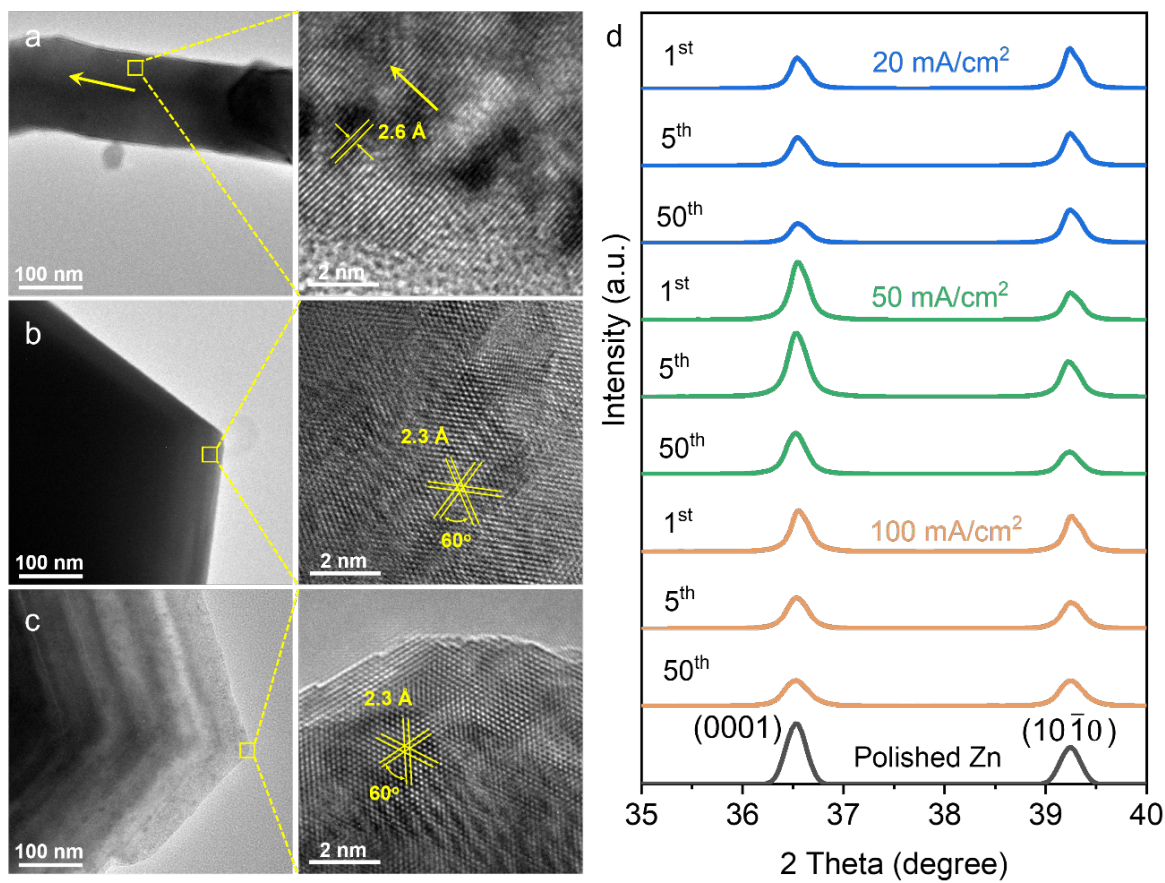

**Figure 2**. TEM and high-resolution TEM images for Zn deposits with the (a) mossy, (b) compact, and (c) dendritic structures obtained at $C_0$ of 0.2 mol/L and a respective $i$ of 20, 50 and 100 mA/cm$^2$ after the 1$^{st}$ cycle. (d) XRD profiles of polished Zn and the cycled Zn anodes under the corresponding cycling conditions.

## II. Growth Mechanisms of the Three Zinc Morphologies

We now focus on the formation mechanisms of the three distinct morphologies. The compact morphology possesses the apparent lowest surface area and the largest fraction of (0001) surface. Given that the surface energy of (0001) is ~ 0.23 J/m$^2$ lower than that of (10$\bar{1}$0) surface according to our density functional theory (DFT) calculations (**Figure S7**), the compact morphology is the most stable one when the deposition is near an equilibrium. Meanwhile, the dendritic morphology is a well-known result[21] of diffusion-limited deposition under high $i$ and/or low $C_0$. However, the thermodynamics does not explain the mossy morphology of a high fraction of (10$\bar{1}$0) surface, which forms at low $i$ and/or high $C_0$, a condition close to an equilibrium.

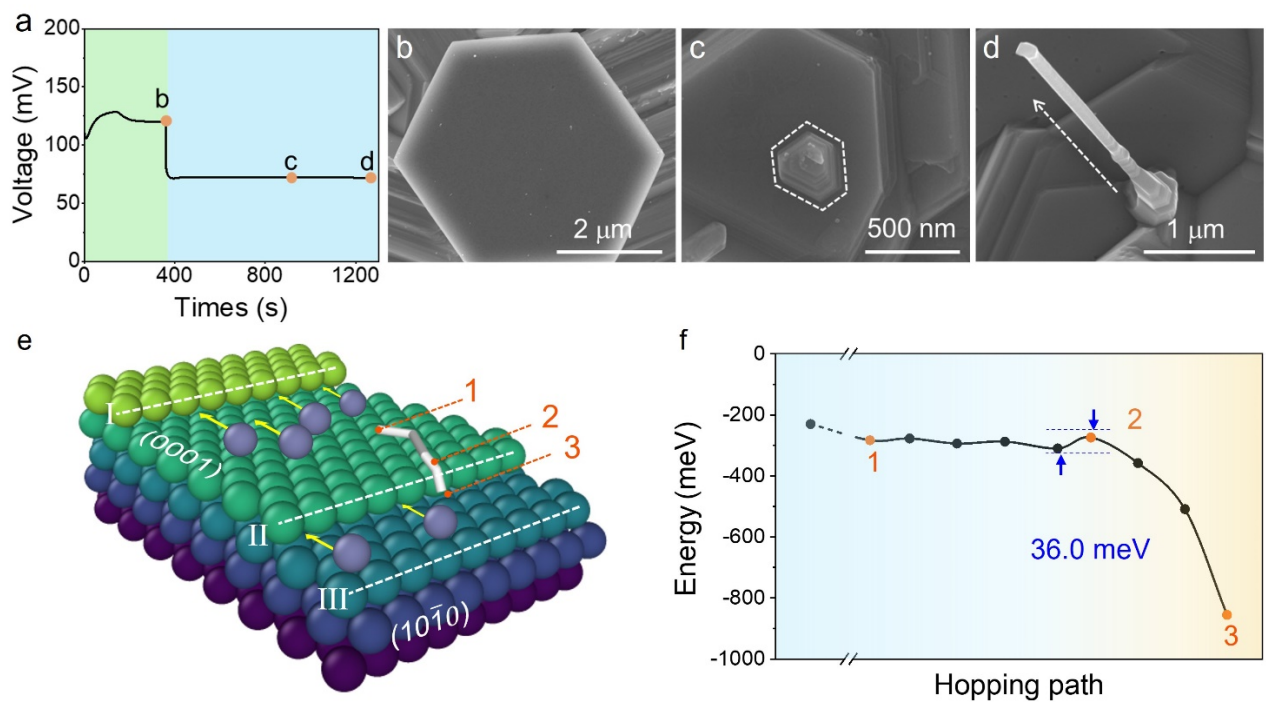

**Figure 3.** Sequential deposition morphologies in a control experiment and proposed mechanism of mossy deposition. (a) Voltage vs. time curve under 50 mA/cm$^2$ (green shaded) and 20 mA/cm$^2$ (blue shaded). (b-c) SEM images for Zn electrodes stopped at corresponding states in (a). (e) An illustration of the step flow process during the spiral growth, in which the terrace between Steps I and II is widening and that between II and III is narrowing. Atoms at different layers of lattice are colored differently, and adatoms are in light purple. (f) The energy profile of the Zn adatom hopping across Step II as shown in (e). The initial, the barrier, and the final positions are marked as 1, 2, and 3 according to (e).

While previous work suggested that the mossy filaments grew from pores of ZnO layer on Zn[13], our control experiment, in which a mossy growth follows a compact deposition, suggests otherwise (**Figure S8 and S9**). In this experiment, we hypothesize according to the oxide mechanism that, the deposition of a compact layer of Zn ($i$ = 50 mA/cm$^2$ and $C_0$ = 0.3 M for six minutes) should largely eliminate sequential mossy deposition (when $i$ is lowered to 20 mA/cm$^2$), because (i) if there were oxide, the formation of compact deposit should signal its complete reduction; and (ii) the oxide should not regrow, as the potential stays below the oxidation potential of Zn. However, in the control experiment, mossy deposition appeared immediately after the initial compact deposition (**Figure 3a-d**), although the population decreased (comparing **Figure S8a5** with **S8b3**). The result suggests that the oxide is at least not the sole reason for the mossy deposition.

As an alternate explanation, we consider surface defects of Zn metal whose existence does not rely on any oxide. The surface defects can provide low coordination surface sites favored by deposition near equilibrium. One such kind of defect is the surface intersection of a screw dislocation. As described by the classic model developed by Burton, Cabrera and Frank[22-23], crystal growth under a low driving force (in our case, low $i$ and high $C_0$) can proceed as a spiral (as seen at the tips of the filaments in **Figure 3c-d** and **Figure S9**) on an otherwise flat surface. While the spiral growth usually leads to pyramids, it can also result in filaments of high aspect ratios if the step edges near the dislocation grows faster than those near the outer edge. Such uncommon growth has been reported for PbS[24], ZnO[25], AlN[26], and Cu[27]. Moreover, a typical dislocation population in Zn[28] could account for the initial number of mossy filaments.

For the growing steps to bunch into a filament, a necessary condition is a low Ehrlich-Schwoebel (ES) barrier, which is supported by our DFT calculations of the ad-atom migration across the $(10\bar{1}0)$ step in the (0001) surface, based on the growth orientation seen in **Figure 3b-c** (see more details in the Supporting Information). An ES barrier is an energy barrier for an adatom to diffuse from an upper terrace to a lower one[29-30]. In a step flow process like the spiral growth, where adatoms go along yellow arrows in **Figure 3e**, a high ES barrier would suppress step bunching and retain equally spaced steps, because any widening terrace will have more arrival atoms incorporated into its step with the upper terrace (Step I in **Figure 3e**). On the contrary, as illustrated in **Figure 3e**, a low ES barrier allow the arrival atoms to move across the step edge (for example, Step II in **Figure 3e**), diminish smaller terraces, and facilitate step bunching. Our DFT calculations indeed support this scenario: a Zn adatom across the step edge of (0001) surface (**Figure 3e**) exhibits an ES barrier of 36.0 meV (**Figure 3f**), which is lower than typical values on metals ($> 100$ meV)[31-32].

Our attribution to surface defects like the screw dislocation suggests that minimizing the associated surface defects may suppress the mossy morphology. In our previous control experiment, we have seen a glimpse of the substrate effect: the mossy deposits grown on a compact Zn morphology exhibits a less population than those grown on the polished Zn foil (**Figure S8**). More effective approaches, which are pursued by others but not yet associated with the above mechanism, include utilizing different polishing methods[33-34] or substrate materials[35-36].

## III. Condensing Electrolytes Inducing Morphological Dynamics

Since the mossy morphology dominates the later stage of cycling (**Figure 1c-e**), it is critical to understand the cause behind. Examination of the electrolytes reveals that the zincate concentration can dramatically increase due to the inefficient charging process. With inductively coupled plasma-optical emission spectrometry (ICP-MS), we measured significant increases in the zincate concentrations in all the cycled electrolytes. **Figure 4a** shows the concentration evolution under four conditions. At $i = 20$ mA/cm$^2$, the zincate concentrations increase by 150% and 65% for $C_0 = 0.2$ and 0.3 M after 50 cycles, respectively, approaching the saturation (~0.5 M) of zincate under room temperature[37]. At $i = 50$ mA/cm$^2$, the concentration increases by 73% and 37% for $C_0 = 0.2$ and 0.3 M after 50 cycles, respectively. Such concentration increase is equivalent to shifting the morphology diagram down along the $y$-axis, as we see in **Figure 1c-e**, and thus triggers the mossy morphology.

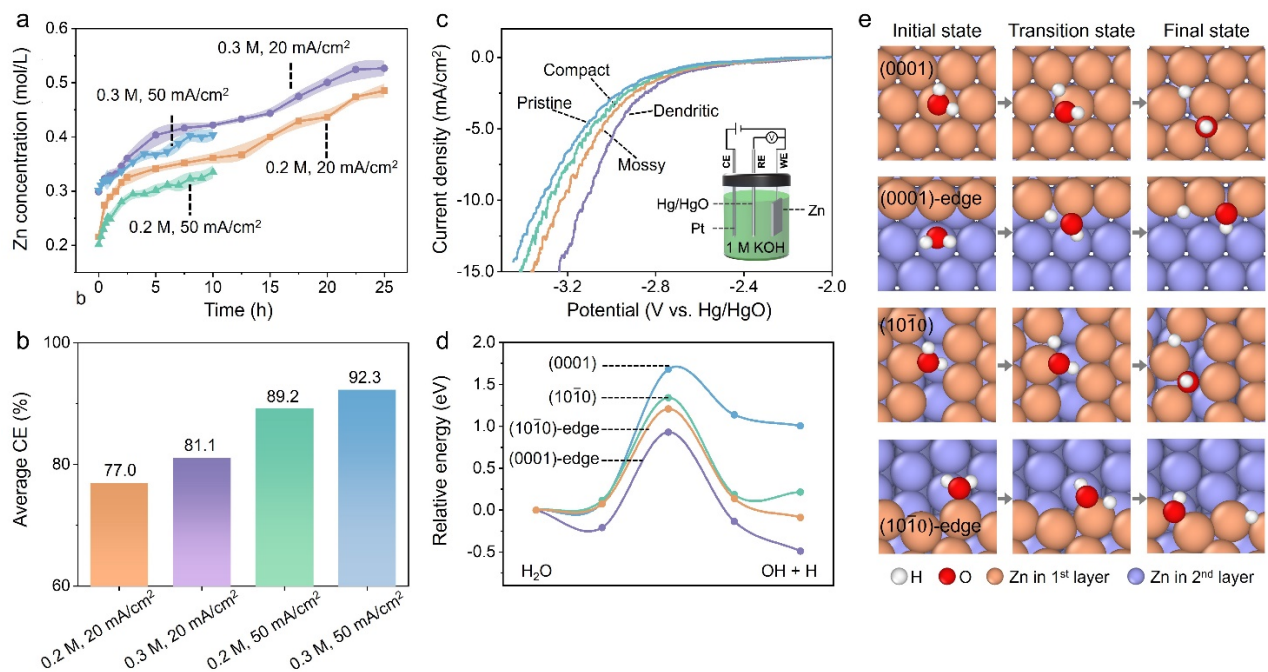

**Figure 4**. Electrolyte compositional evolution. (a) Zincate concentrations analyzed at the ends of charging steps. The shaded areas indicate the errors. (b) Average CE calculated for the four conditions analyzed in (a). (c) LSV curves for Zn with different morphologies; inset displays the setup of the experiment. The supporting electrolyte is 1 M KOH solution. (d) Reaction energy diagram of water dissociation on Zn(0001), Zn(10$\bar{1}$0), Zn (0001) edge, and Zn(10$\bar{1}$0) edge, and (e) structures of the corresponding initial state, transition state, and final state.

Besides changing the zinc morphology, the condensing zincate electrolyte also induces the formation of ZnO precipitation on the air cathode. As shown in **Figure S10**, more and more ZnO are found on the cycled cathode, as the cycling proceeds. The precipitation is more severe on the cathode than the rest of the battery, because the concentration of $Zn(OH)_4^{2-}$ ions is higher near the cathode due to a favorable electrostatic potential and the oxygen evolution reaction during charging consumes the $OH^-$, both facilitating the precipitation reaction of $Zn(OH)_4^{2-} \rightarrow ZnO + H_2O + 2OH^-$. These precipitates were found to passivate the catalytic sites in cathodes[38].

Underlying the increasing zincate concentration is a heavy HER, which is thermodynamically favored over Zn deposition during charging[12]. Water loss, another possible cause, is ruled out given the stable potassium concentrations in all the cases (**Figure S11**). Since HER consumes the charge input, while no obvious side-reactions exist during the discharging process, a net increase of zincate concentration occurs under the fixed-capacity cycling protocol. Such net increase corresponds to a loss of Coulombic efficiency (CE). Calculations based on the net increase of zincate concentration in **Figure 4a** and Faraday's law lead to an average CE ranging from 77% to 92% for the four cases (**Figure 4b**). Although we are always tempted to report the values close to 100% calculated directly from the cycling data (e.g., **Figure S1b-d**), they are the artefacts of the constant-capacity cycling protocol.

## IV. Impacts of Zn Morphology on HER

A counterintuitive observation from **Figure 4b** is that the CE is much lower at low $i$ under the same $C_0$, as we would expect the larger driving force at high $i$ would result in a heavier HER and thus lower CE[39-40]. The cause goes back to the morphological difference: the compact deposit formed under higher $i$ is lower in the specific area than the mossy deposit formed at low $i$. Moreover, the extent of HER depends on not only the surface area but also the catalytic activity, which can differ substantially among the three types of morphologies. To examine the catalytic differences toward HER, we performed linear sweep voltammetry (LSV, **Figure 4c**) in 1 M KOH on the three morphologies (attained under $C_0 = 0.2$ M and $i = 20$, 50, or 100 mA/cm$^2$, respectively). We first estimate electrochemically active surface area (ECSA) through the double-layer capacitance (**Figure S12**)[41-42]. Assuming a specific capacitance of 0.31 mF/cm$^2$ for Zn[43], the ESCA is about 4.04, 3.72, 3.38, and 3.31 cm$^2$ for the dendritic, mossy, compact, and pristine morphologies, respectively, as we would expect. The measured current is then normalized by ECSA and plotted in **Figure 4c**, which shows that the dendritic and the mossy morphologies are more catalytic towards HER than the compact and pristine ones.

To understand the origin of the different catalytic activities, we carry out transition state calculations based on the DFT. The rate-determining step of HER in alkaline electrolyte is believed to be the dissociation of water molecule[44-46], which supplies hydrogen atoms for subsequent hydrogen formation.

Given that the compact structure exposes majorly the (0001) facets, the mossy structures expose majorly the (10$\bar{1}$0) facets and their edges, and the dendritic structures expose lots of (0001) facet edges (**Figure S6c**), we examine the water dissociation on the (0001) and (10$\bar{1}$0) surfaces, together with their edges. **Figure 4d** shows that the energy barrier to break the O-H bond in a water molecule is 1.68 eV on the (0001) surface, 1.34 eV on the (10$\bar{1}$0) surface, 1.21 eV at the (10$\bar{1}$0) surface edge, and 0.93 eV at the (0001) surface edge. The corresponding structures of the critical states can be found in **Figure 4e**. Therefore, our DFT results confirm that the dendritic structures exhibit the most severe HER, followed by the mossy structures, and then the compact structure, explaining the different HER activities in **Figure 4c** and the increasing trends of zincate concentration in **Figure 4a**.

## V. Strategy to Improve the Battery Performances

Inspired by the discovery that the HER-induced CE loss results in the increasing zincate concentration under the fixed-capacity cycling protocol and subsequently lead to the irreversible changes of electrode morphologies, we propose an overcharge-cycling protocol: a battery is intentionally overcharged to compensate the CE loss by HER. In theory, this strategy could maintain a stable and unsaturated zincate concentration, which help retain the compact morphology of the anode and avoid the oxide passivation on cathode. **Figure 5a-b** illustrates the battery performances with different overcharge ratios and demonstrate that the battery with a 15% overcharging exhibits the longest cycles and the greatest total delivered discharge capacity, which are up to 2.2 times of those without overcharging. The failure of the cells likely root in the cathode sides based on postmortem characterizations. For the ones with no or low overcharging, the cathodes are full of ZnO precipitates (**Figure S10**), which severely passivate the catalysts. For the ones with high overcharging, carbon corrosion during the oxygen evolution reaction seems more severe to undermine the cathode stability. For the moderate 15% overcharging, the cathode remains the most stable, so does the zincate concentration as inferred from the significantly less mossy Zn in **Figure 5c-d**. This success suggests that a more delicate control of the zincate concentration would lead to even better battery performances.

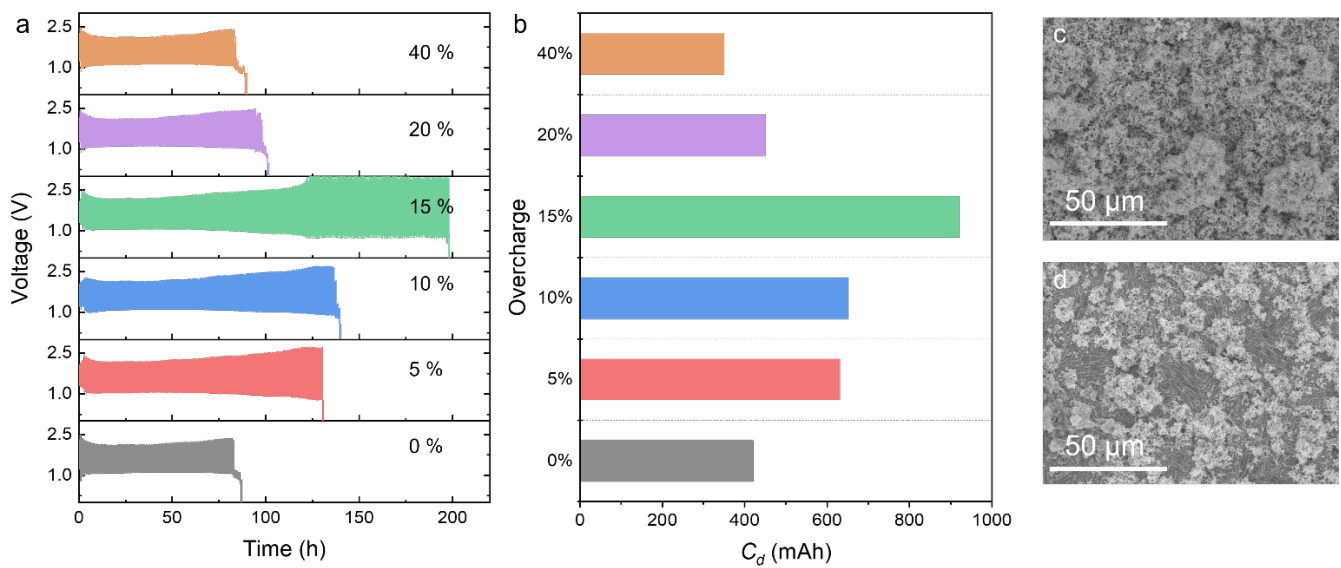

**Figure 5**. (a) Representative galvanostatic discharging and charging curves with a 0%, 5%, 10%, 15%, 20%, and 40% overcharging ratio under $C_0 = 0.2$ M, $i = 10$ mA/cm$^2$, and a discharging capacity of 5 mAh/cm$^2$. (b) The total delivered discharging capacities ($C_d$) before failure for cells in (a). SEM images of Zn anodes with (c) 0 % and (d) 15 % overcharging ratio after the 50$^{th}$ charging cycle.

**Conclusion**

In summary, we reveal the morphological dynamics of electrodes in alkaline RZABs and uncover their intimate relationship with electrolyte evolution and HER. Overall, the noticeable CE loss by HER increases the zincate concentration, which increases the mossy deposition and induces the oxide passivation. The mossy deposition in turn aggravates HER due to its catalytic effect and accelerates the concentration increase. All the entangled issues finally lead to the short cycling lives of RZABs. We show the need for more rigorous tests of the true CE efficiencies and propose an overcharge-cycling protocol to significantly boost the reversibility of alkaline electrolyte and electrode morphologies.

**Acknowledgements**

This work was financially supported by the Guangdong Provincial Key Laboratory of Computational Science and Material Design (Grant No. 2019B030301001), the Introduced Innovative R&D Team of Guangdong (Grant No. 2017ZT07C062), the Shenzhen Science and Technology Innovation Commission (No. JCYJ20200109141412308), the National Foundation of Natural Science, China (No. 52022002), and the Research Grant Council, Hong Kong (No. C1002-21G). The calculations were carried out on the Taiyi cluster supported by the Center for Computational Science and Engineering of Southern University of Science and Technology and also on The Major Science and Technology Infrastructure Project of Material Genome Big-science Facilities Platform supported by Municipal Development and Reform Commission of Shenzhen.

**Reference**

1. Arnot, D. J.; Lim, M. B.; Bell, N. S.; Schorr, N. B.; Hill, R. C.; Meyer, A.; Cheng, Y.-T.; Lambert, T. N., High Depth-of-Discharge Zinc Rechargeability Enabled by a Self-Assembled Polymeric Coating. *Adv. Energy Mater.* **2021**, *11*, 2101594.

2. Xie, C., et al., Discovering the Intrinsic Causes of Dendrite Formation in Zinc Metal Anodes: Lattice Defects and Residual Stress. *Angew. Chem. Int. Ed.* **2023**, *62*, e202218612.

3. Chen, S.; Wang, T.; Ma, L.; Zhou, B.; Wu, J.; Zhu, D.; Li, Y. Y.; Fan, J.; Zhi, C., Aqueous Rechargeable Zinc Air Batteries Operated at −110°C. *Chem* **2023**, *9*, 497-510.

4. Jin, S.; Chen, P.-Y.; Qiu, Y.; Zhang, Z.; Hong, S.; Joo, Y. L.; Yang, R.; Archer, L. A., Zwitterionic Polymer Gradient Interphases for Reversible Zinc Electrochemistry in Aqueous Alkaline Electrolytes. *J. Am. Chem. Soc.* **2022**, *144*, 19344-19352.

5. Liu, J.-N.; Zhao, C.-X.; Wang, J.; Ren, D.; Li, B.-Q.; Zhang, Q., A Brief History of Zinc–Air Batteries: 140 Years of Epic Adventures. *Energy Environ. Sci.* **2022**, *15*, 4542-4553.

6. Yuan, Z.; Yin, Y.; Xie, C.; Zhang, H.; Yao, Y.; Li, X., Advanced Materials for Zinc-Based Flow Battery: Development and Challenge. *Adv. Mater.* **2019**, *31*, 1902025.

7. Shinde, S. S.; Jung, J. Y.; Wagh, N. K.; Lee, C. H.; Kim, D.-H.; Kim, S.-H.; Lee, S. U.; Lee, J.-H., Ampere-Hour-Scale Zinc–Air Pouch Cells. *Nat. Energy* **2021**, *6*, 592-604.

8. Luo, F.; Zhu, J.; Ma, S.; Li, M.; Xu, R.; Zhang, Q.; Yang, Z.; Qu, K.; Cai, W.; Chen, Z., Regulated Coordination Environment of Ni Single Atom Catalyst toward High-Efficiency Oxygen Electrocatalysis for Rechargeable Zinc-Air Batteries. *Energy Storage Mater.* **2021**, *35*, 723-730.

9. Zhang, W., et al., Surface Oxygenation Induced Strong Interaction between Pd Catalyst and Functional Support for Zinc–Air Batteries. *Energy Environ. Sci.* **2022**, *15*, 1573-1584.

10. Wang, R., et al., Precise Identification of Active Sites of a High Bifunctional Performance 3d Co/N-C Catalyst in Zinc-Air Batteries. *Chemical Engineering Journal* **2022**, *433*, 134500.

11. Tang, K.; Hu, H.; Xiong, Y.; Chen, L.; Zhang, J.; Yuan, C.; Wu, M., Hydrophobization Engineering of the Air–Cathode Catalyst for Improved Oxygen Diffusion Towards Efficient Zinc–Air Batteries. *Angew. Chem. Int. Ed.* **2022**, *61*, e202202671.

12. Fu, J.; Cano, Z. P.; Park, M. G.; Yu, A.; Fowler, M.; Chen, Z., Electrically Rechargeable Zinc–Air Batteries: Progress, Challenges, and Perspectives. *Adv. Mater.* **2017**, *29*, 1604685.

13. Desai, D.; Wei, X.; Steingart, D. A.; Banerjee, S., Electrodeposition of Preferentially Oriented Zinc for Flow-Assisted Alkaline Batteries. *J. Power Sources* **2014**, *256*, 145-152.

14. Naybour, R. D., Morphologies of Zinc Electrodeposited from Zinc-Saturated Aqueous Alkaline Solution. *Electrochim. Acta* **1968**, *13*, 763-769.

15. Wang, R. Y.; Kirk, D. W.; Zhang, G. X., Effects of Deposition Conditions on the Morphology of Zinc Deposits from Alkaline Zincate Solutions. *J. Electrochem. Soc.* **2006**, *153*, C357.

16. Wang, R. Y.; Kirk, D. W.; Zhang, G. X., Characterization and Growth Mechanism of Filamentous Zinc Electrodeposits. *ECS Trans.* **2007**, *2*, 19.

17. Zheng, J.; Archer Lynden, A., Controlling Electrochemical Growth of Metallic Zinc Electrodes: Toward Affordable Rechargeable Energy Storage Systems. *Sci. Adv.*, *7*, eabe0219.

18. Lee, H.-J.; Lim, J.-M.; Kim, H.-W.; Jeong, S.-H.; Eom, S.-W.; Hong, Young T.; Lee, S.-Y., Electrospun Polyetherimide Nanofiber Mat-Reinforced, Permselective Polyvinyl Alcohol Composite Separator Membranes: A Membrane-Driven Step Closer toward Rechargeable Zinc–Air Batteries. *J. Membr. Sci.* **2016**, *499*, 526-537.

19. Yang, Y.; Yang, H.; Zhu, R.; Zhou, H., High Reversibility at High Current Density: The Zinc Electrodeposition Principle Behind the "Trick". *Energy Environ. Sci.* **2023**,16, 2723-2731.

20. Bockris, J. O. M.; Nagy, Z.; Drazic, D., On the Morphology of Zinc Electrodeposition from Alkaline Solutions. *J. Electrochem. Soc.* **1973**, *120*, 30.

21. Vishnugopi, B. S.; Hao, F.; Verma, A.; Mukherjee, P. P., Surface Diffusion Manifestation in Electrodeposition of Metal Anodes. *Phys. Chem. Chem. Phys.* **2020**, *22*, 11286-11295.

22. Burton, W. K.; Cabrera, N.; Frank, F. C., Role of Dislocations in Crystal Growth. *Nature* **1949**, *163*, 398-399.

23. Burton, W. K.; Cabrera, N.; Frank, F. C.; Mott, N. F., The Growth of Crystals and the Equilibrium Structure of Their Surfaces. *Philos. Trans. R. Soc. A* **1997**, *243*, 299-358.

24. Lau, Y. K. A.; Chernak, D. J.; Bierman, M. J.; Jin, S., Formation of Pbs Nanowire Pine Trees Driven by Screw Dislocations. *J. Am. Chem. Soc.* **2009**, *131*, 16461-16471.

25. Morin, S. A.; Bierman, M. J.; Tong, J.; Jin, S., Mechanism and Kinetics of Spontaneous Nanotube Growth Driven by Screw Dislocations. *Science* **2010**, *328*, 476-480.

26. Meng, F.; Estruga, M.; Forticaux, A.; Morin, S. A.; Wu, Q.; Hu, Z.; Jin, S., Formation of Stacking Faults and the Screw Dislocation-Driven Growth: A Case Study of Aluminum Nitride Nanowires. *ACS Nano* **2013**, *7*, 11369-11378.

27. Meng, F.; Jin, S., The Solution Growth of Copper Nanowires and Nanotubes Is Driven by Screw Dislocations. *Nano Lett.* **2012**, *12*, 234-239.

28. Berghezan, A.; Fourdeux, A.; Amelinckx, S., Transmission Electron Microscopy Studies of Dislocations and Stacking Faults in a Hexagonal Metal: Zinc. *Acta Metall.* **1961**, *9*, 464-490.

29. Schwoebel, R. L.; Shipsey, E. J., Step Motion on Crystal Surfaces. *J. Appl. Phys.* **1966**, *37*, 3682-3686.

30. Ehrlich, G.; Hudda, F. G., Atomic View of Surface Self‐Diffusion: Tungsten on Tungsten. *J. Chem. Phys.* **1966**, *44*, 1039-1049.

31. Xiang, S. K.; Huang, H., Ab Initio Determination of Ehrlich–Schwoebel Barriers on Cu{111}. *Appl. Phys. Lett.* **2008**, *92*, 101923.

32. Vrijmoeth, J.; van der Vegt, H. A.; Meyer, J. A.; Vlieg, E.; Behm, R. J., Surfactant-Induced Layer-by-Layer Growth of Ag on Ag(111): Origins and Side Effects. *Phys. Rev. Lett.* **1994**, *72*, 3843-3846.

33. He, P.; Huang, J., Detrimental Effects of Surface Imperfections and Unpolished Edges on the Cycling Stability of a Zinc Foil Anode. *ACS Energy Lett.* **2021**, *6*, 1990-1995.

34. Zhang, Z.; Said, S.; Smith, K.; Zhang, Y. S.; He, G.; Jervis, R.; Shearing, P. R.; Miller, T. S.; Brett, D. J. L., Dendrite Suppression by Anode Polishing in Zinc-Ion Batteries. *J. Mater. Chem. A* **2021**, *9*, 15355-15362.

35. Pu, S. D., et al., Achieving Ultrahigh-Rate Planar and Dendrite-Free Zinc Electroplating for Aqueous Zinc Battery Anodes. *Adv. Mater.* **2022**, *34*, 2202552.

36. Qian, G., et al., Structural, Dynamic, and Chemical Complexities in Zinc Anode of an Operating Aqueous Zn-Ion Battery. *Adv. Energy Mater.* **2022**, *12*, 2200255.

37. Dirkse, T. P., Aqueous Potassium Hydroxide as Electrolyte for the Zinc Electrode. *J. Electrochem. Soc.* **1987**, *134*, 11-13.

38. Kim, H.-W.; Lim, J.-M.; Lee, H.-J.; Eom, S.-W.; Hong, Y. T.; Lee, S.-Y., Artificially Engineered, Bicontinuous Anion-Conducting/-Repelling Polymeric Phases as a Selective Ion Transport Channel for Rechargeable Zinc–Air Battery Separator Membranes. *J. Mater. Chem. A* **2016**, *4*, 3711-3720.

39. Luo, Y.; Tang, L.; Khan, U.; Yu, Q.; Cheng, H.-M.; Zou, X.; Liu, B., Morphology and Surface Chemistry Engineering toward Ph-Universal Catalysts for Hydrogen Evolution at High Current Density. *Nat. Commun.* **2019**, *10*, 269.

40. Luo, Y.; Zhang, Z.; Yang, F.; Li, J.; Liu, Z.; Ren, W.; Zhang, S.; Liu, B., Stabilized Hydroxide-Mediated Nickel-Based Electrocatalysts for High-Current-Density Hydrogen Evolution in Alkaline Media. *Energy Environ. Sci.* **2021**, *14*, 4610-4619.

41. McCrory, C. C. L.; Jung, S.; Peters, J. C.; Jaramillo, T. F., Benchmarking Heterogeneous Electrocatalysts for the Oxygen Evolution Reaction. *J. Am. Chem. Soc.* **2013**, *135*, 16977-16987.

42. Lukowski, M. A.; Daniel, A. S.; Meng, F.; Forticaux, A.; Li, L.; Jin, S., Enhanced Hydrogen Evolution Catalysis from Chemically Exfoliated Metallic Mos2 Nanosheets. *J. Am. Chem. Soc.* **2013**, *135*, 10274-10277.

43. Guo, M.; Li, X.; Huang, Y.; Li, L.; Li, J.; Lu, Y.; Xu, Y.; Zhang, L., Co2-Induced Fibrous Zn Catalyst Promotes Electrochemical Reduction of Co2 to Co. *Catalysts* **2021**, *11*, 477.

44. Durst, J.; Siebel, A.; Simon, C.; Hasché, F.; Herranz, J.; Gasteiger, H. A., New Insights into the Electrochemical Hydrogen Oxidation and Evolution Reaction Mechanism. *Energy Environ. Sci.* **2014**, *7*, 2255-2260.

45. Subbaraman, R.; Tripkovic, D.; Strmcnik, D.; Chang, K.-C.; Uchimura, M.; Paulikas Arvydas, P.; Stamenkovic, V.; Markovic Nenad, M., Enhancing Hydrogen Evolution Activity in Water Splitting by Tailoring Li+-Ni(Oh)2-Pt Interfaces. *Science* **2011**, *334*, 1256-1260.

46. Wang, P.; Zhang, X.; Zhang, J.; Wan, S.; Guo, S.; Lu, G.; Yao, J.; Huang, X., Precise Tuning in Platinum-Nickel/Nickel Sulfide Interface Nanowires for Synergistic Hydrogen Evolution Catalysis. *Nat. Commun.* **2017**, *8*, 14580.

**TOC Graphic**

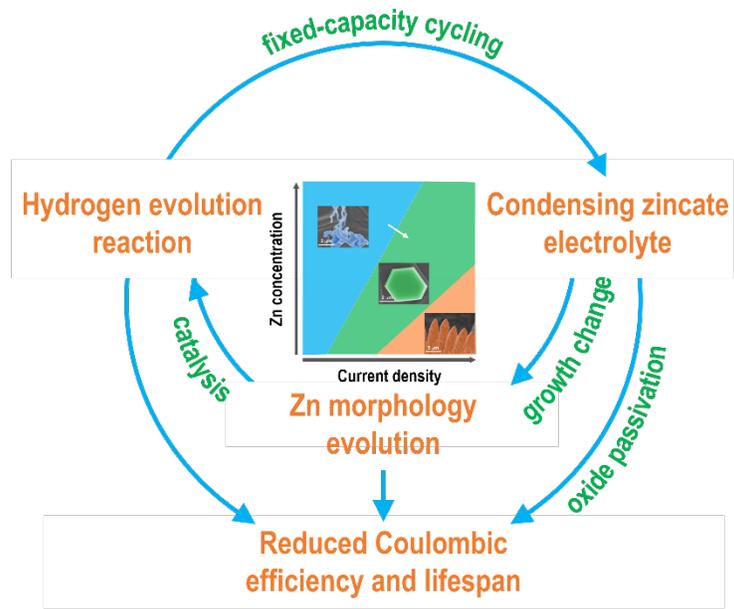

# Supporting Information of "Decoding the Mechanisms of Reversibility Loss in Rechargeable Zinc-Air Batteries"

Zhibin Yi[1,2], Liangyu Li[2], Cheuk Kai Chan,[3] Yaxin Tang[1], Zhouguang Lu[1], Chunyi Zhi,[4] Qing Chen[2,3]*, and Guangfu Luo[1,5]*

[1]Department of Materials Science and Engineering, Southern University of Science and Technology, Shenzhen 518055, P. R. China

[2]Department of Mechanical and Aerospace Engineering, The Hong Kong University of Science and Technology, Clear Water Bay, Hong Kong, P. R. China

[3]Department of Chemistry, The Hong Kong University of Science and Technology, Clear Water Bay, Hong Kong, P. R. China

[4]Department of Materials Science and Engineering, City University of Hong Kong, Kowloon, Hong Kong, China.

[5]Guangdong Provincial Key Laboratory of Computational Science and Material Design, Southern University of Science and Technology, Shenzhen 518055, China

*E-mail: chenqing@ust.hk, luogf@sustech.edu.cn

**Experimental and Computational Details**

**Battery Assembly and Testing:** The battery module is a stack-type cell, as shown in Figure S1a. Distance between the anode and cathode surfaces is about 1.2 cm, the electrolyte volume is 4 mL, and the reactive surface for Zn anode is 3 cm$^2$. A specific amount of ZnO powder is dissolved into a 6 M KOH solution to prepare the electrolyte with corresponding zincate concentration. The catalyst slurry consists of an equal weight of Pt/C (20 wt% loading, Sigma-Aldrich) and IrO$_2$ (Sigma-Aldrich) powder, isopropyl alcohol and deionized water, and 0.5% Nafion solution with a volume ratio of 7 : 2.5 : 0.5 (20 mg of active catalyst and 2 mL of total liquid volume). The suspension is dispersed on hydrophobic carbon paper via a drop-casting method and dried at 60 ºC for 24 h. The catalyst loading is about 2 mg/cm$^2$. Galvanostatic cycling tests are performed with a Neware battery testing station (CT-4008) and the same discharging-charging capacity of 5 mAh/cm$^2$ is applied for all experiments unless stated otherwise.

The HER performance tests of different Zn morphologies are carried out at room temperature, and a three-electrode system is used in an electrochemical workstation (Biologic). N$_2$-saturated 1 M KOH solution is used as electrolyte and is changed before each test. Pt mesh and Hg/HgO electrode are used as the counter and reference electrode, respectively. The linear sweep voltammetry (LSV) is conducted at a scanning rate of 1 mV/s. Steady-state CV curves are recorded at a series of scan rates of 20, 40, 60, 80, and 100 mV s$^{-1}$.

**Characterization:** Surface morphology and composition of the cycled Zn anodes are characterized by SEM (Hitachi Regulus8230) and EDS (Bruker XFlash7). The detailed microstructures of the Zn deposits are observed on a field emission TEM (Tecnai G2 F30 U-TWIN). XRD analysis are performed using Bruker-AXS Microdiffractometer (D8 Advance X-ray diffractometer, Cu Kα, λ = 1.54056 Å, 40 kV and 40 mA). Anodes obtained from reacted cells are rinsed in 80 mL deionized water for seven times to remove residual salts and then dried and stored under vacuum before further characterization. To analyze the compositional change of the cycled electrolyte solutions, electrolyte aliquots are taken out from cells at different stages of cycling under four different testing conditions ($C_0$ = 0.2 or 0.3 M; $i$ = 20 or 50 mA/cm$^2$) and tested by the inductively coupled plasma mass spectrometry (ICP-MS).

**Computations**: The density functional theory (DFT) calculations are performed using the Vienna *Ab-initio* Simulation Package (VASP)[1-3] with the Perdew-Burke-Ernzerhof functional[2]. The van der Waals interactions are included using the DFT-D3 approach.[4] The projector augmented wave method is used with the pseudopotential Zn(3d$^{10}$4s$^2$) for zinc, H(1s$^1$) for hydrogen, and O(2s$^2$2p$^4$) for oxygen. A plane-wave energy cutoff of 393 eV and a *k*-spacing of 2π/40 Å$^{-1}$ is used in each periodic direction. The total energy and force convergence are 10$^{-5}$ eV and 0.01 eV/Å, respectively.

The Zn surface energy $\sigma$ is calculated according to Eqn. 1,

$$\sigma = (E_{slab} - N*E_{bulk})/2A, \qquad (1)$$

where $E_{slab}$, $N$, and $A$ is the total energy, number of Zn atom, and total surface area of the slab model; $E_{bulk}$ is the energy of Zn bulk per atom. The supercell slabs for the (0001) and (10$\bar{1}$0) surface possesses sizes of 2.7 × 2.7 × 15.2 Å$^3$ and 2.7 × 4.9 × 10.6 Å$^3$, respectively, with three middle layers fixed to their positions in the bulk. An additional vacuum layer of 12 Å is added in the out-of-plane direction.

The adsorption energy for Zn adatom around the step-edge is defined as Eqn. 2,

$$E_{ad} = E_{adsorbed} - E_{clean} - E_{atom}, \qquad (2)$$

where $E_{adsorbed}$ and $E_{clean}$ are the total energies of surfaces with and without Zn adatom, respectively, and $E_{atom}$ is the total energy of an isolated Zn atom. The supercell size is 25 × 16 × 30 Å$^3$, with four bottom layers fixed to their positions in the bulk. An additional vacuum layer of 12 Å is added in the out-of-plane directions. The activation barriers are calculated using the climbing nudged elastic band method[5].

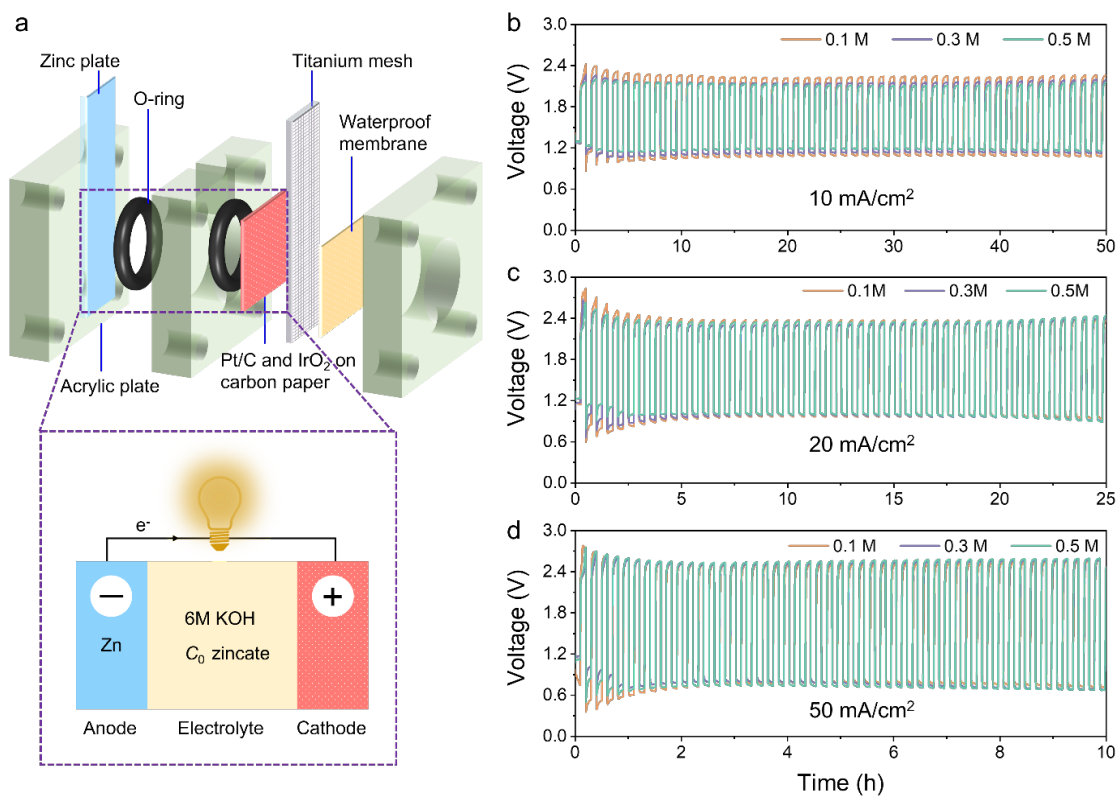

**Figure S1.** (a) Schematic of the RZAB cell employed in this study. (b–c) Representative galvanostatic discharging and charging curves at $i$ = 10, 20, and 50 mA/cm$^2$, respectively, with different initial zincate concentration $C_0$. The relatively large voltage differences between charging and discharging at the higher currents are due to limited attention given to the cathode catalysts.

**Table S1**. Electrolyte, current density, and areal capacity of stack-cell of RZABs from literature.

| Electrolyte | Current density (mA/cm$^2$) | Areal capacity (mAh/cm$^2$) | Ref. |
|---|---|---|---|
| 3.2 M KOH, 1.5 M KF, sat. ZnO | 6 (discharging) 3 (charging) | - | 6 |
| 6 M KOH | 20 | 1.7 | 7 |
| 6 M KOH | 8 | 10 | 8 |
| 6 M KOH, 0.2 M Zn(CH$_3$COO)$_2$ | 10 | 1.7 | 9 |
| 6 M KOH, 0.2 M Zn(CH$_3$COO)$_2$ | 2 | 0.17 | 10 |
| 6 M KOH, 0.2 M Zn(CH$_3$COO)$_2$ | 10 | 0.85 | 11 |
| 6 M KOH, 0.2 M Zn(CH$_3$COO)$_2$ | 25 | 0.85 | 12 |
| 6 M KOH, 0.2 M ZnCl$_2$ | 10 | 10 | 13 |
| 6 M KOH, 0.2 M ZnCl$_2$ | 10 | 1.7 | 14 |
| 6 M KOH, 0.4 M ZnO | 25 (discharging) 15 (charging) | 2.1; 1.25 | 15 |
| 8 M KOH | 25 | - | 16 |

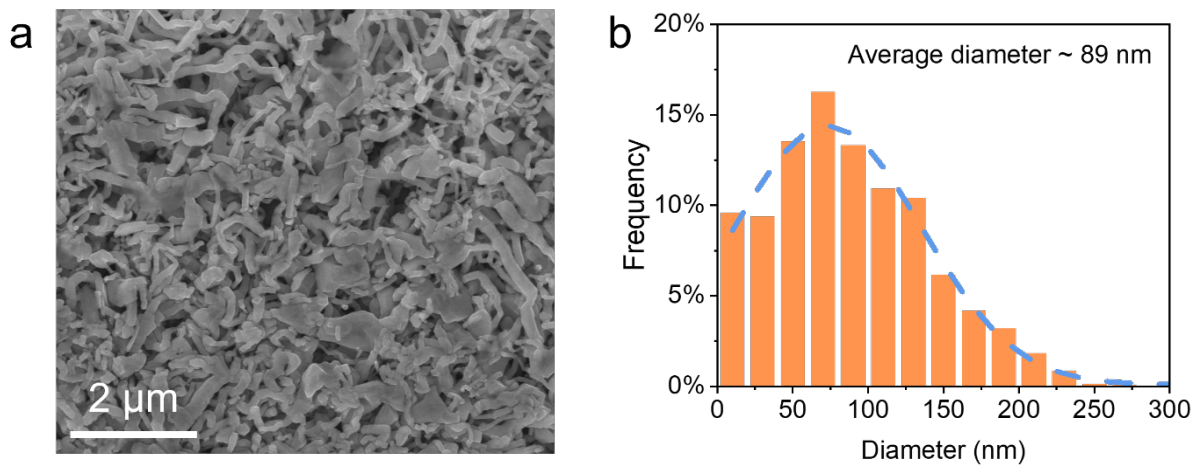

**Figure S2.** (a) SEM image of mossy zinc deposit after the 1st discharging-charging cycle under $C_0 = 0.3$ M and $i = 20$ mA/cm². (b) Diameter distribution of the zinc filaments in (a).

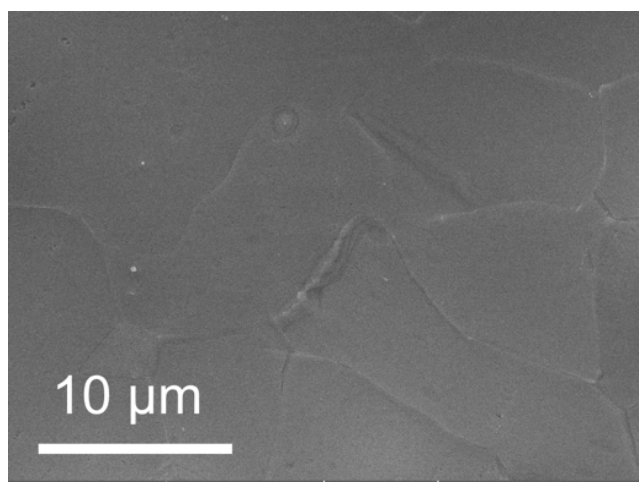

**Figure S3.** SEM image of polished Zn plate

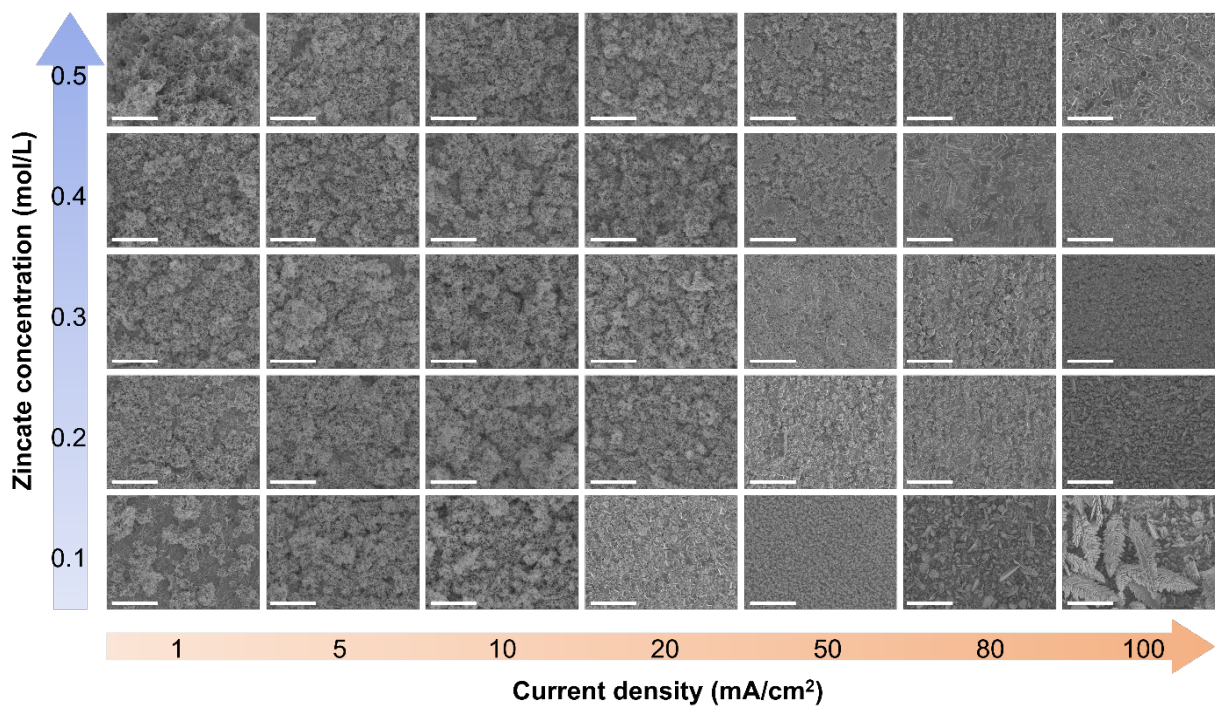

**Figure S4**. SEM images of the Zn anodes after the 1st discharging-charging cycle under the corresponding current densities and initial zincate concentrations. The white bars indicate 50 μm.

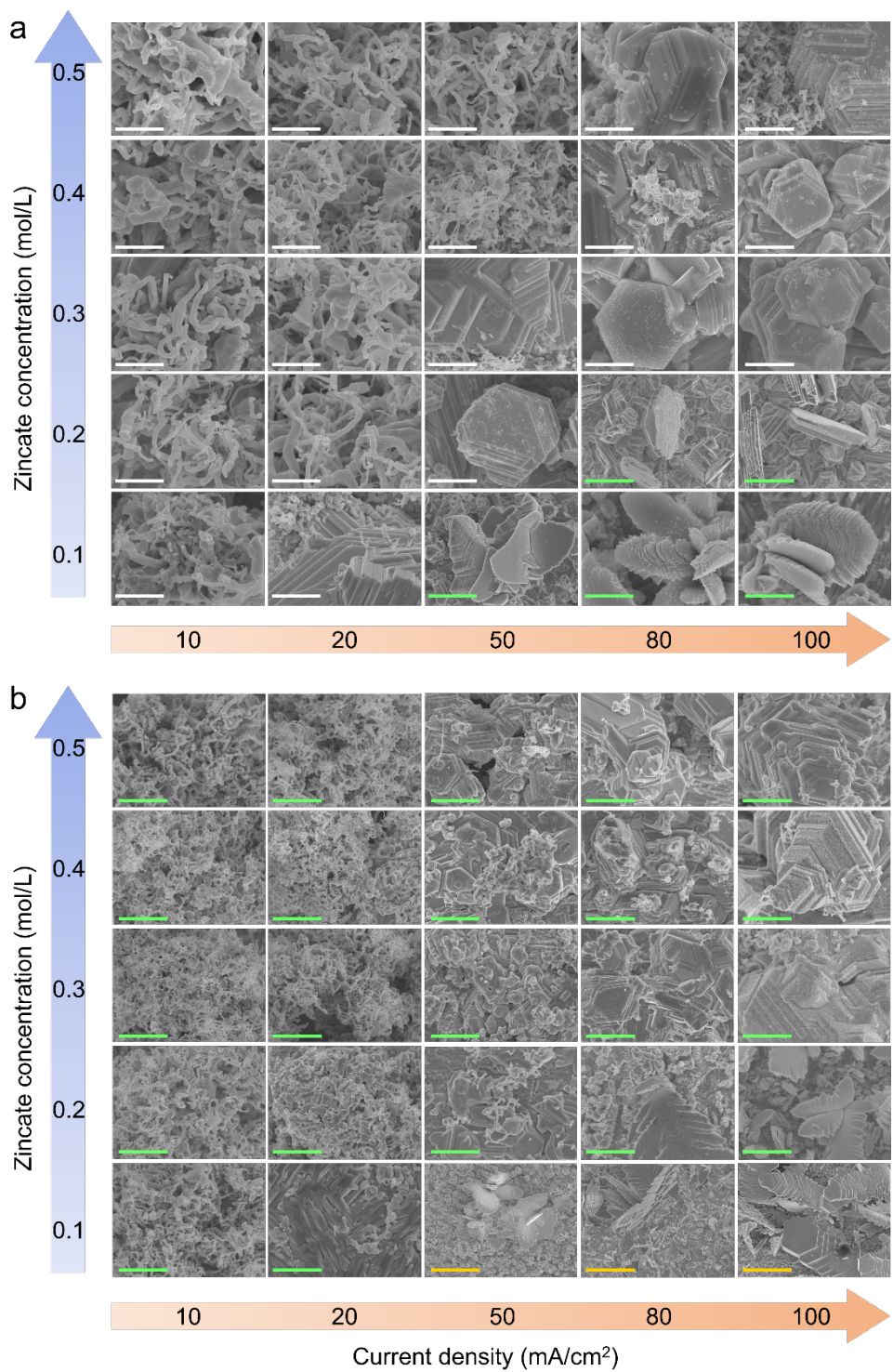

**Figure S5**. SEM images of zinc anodes after the (a) 5th and (b) 50th charging cycle under different conditions. The white, green, and yellow bars indicate 2, 10, and 50 μm, respectively.

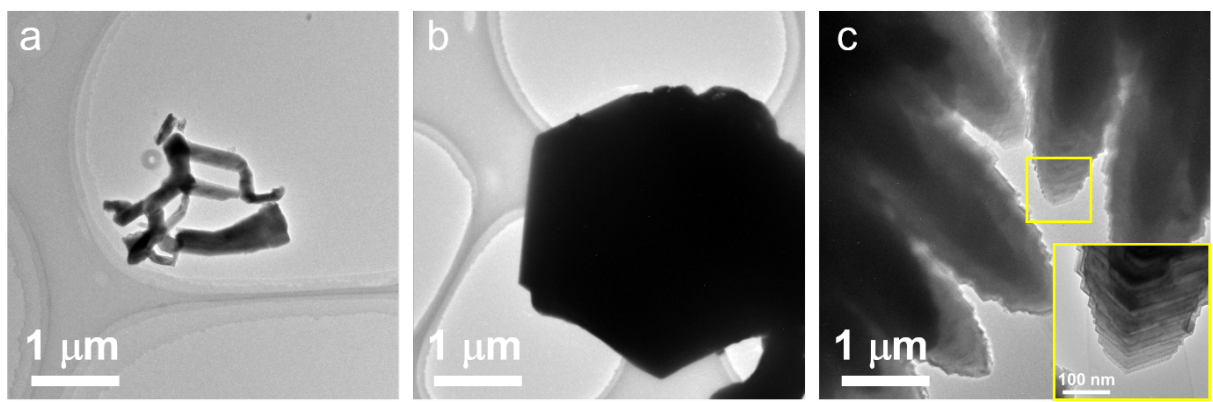

**Figure S6**. TEM image of (a) mossy, (b) compact and (c) dendritic Zn structures. Inset in panel (c) shows a magnified area as labeled in the yellow square.

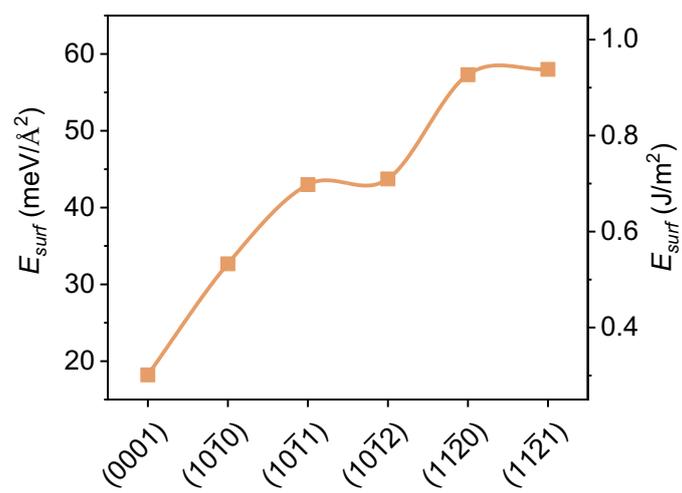

**Figure S7.** Predicted surface energies for various crystal planes of Zn metal.

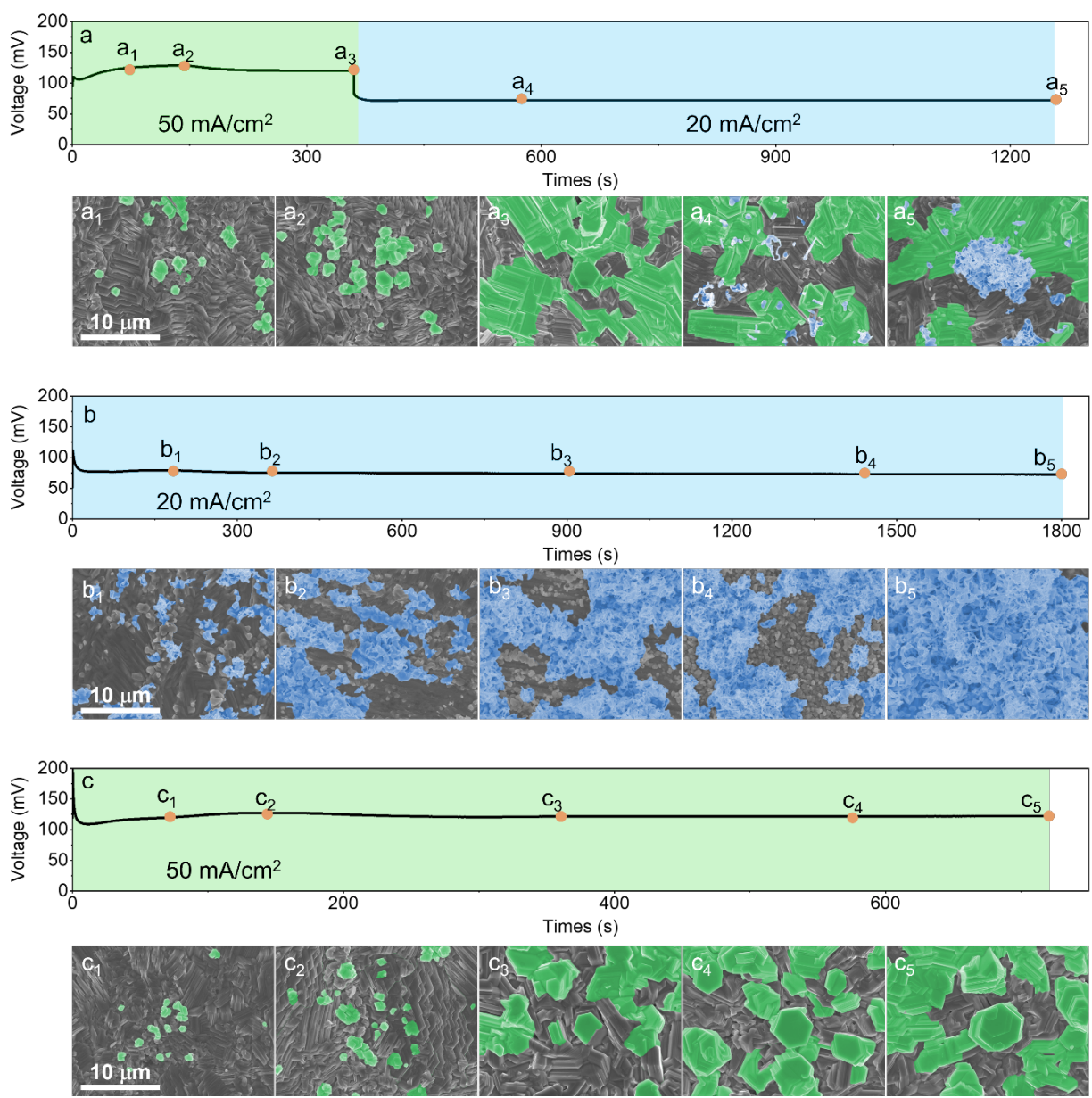

**Figure S8.** Time dependence of battery cell voltage under $C_0$ = 0.3 M for (a) deposition at 50 mA/cm² with a capacity of ($a_1$) 1, ($a_2$) 2, and ($a_3$) 5 mAh/cm², followed by an additional deposition at 20 mA/cm² with a capacity of ($a_4$) 3 and ($a_5$) 5 mAh/cm², (b) deposition at 20 mA/cm² with a capacity of ($b_1$) 1, ($b_2$) 2, ($b_3$) 5, ($b_4$) 8, and ($b_5$) 10 mAh/cm², and (c) deposition at 50 mA/cm² with a capacity of ($c_1$) 1, ($c_2$) 2, ($c_3$) 5, ($c_4$) 8, and ($c_5$) 10 mAh/cm². Below each curve are SEM images of the Zn anodes at the corresponding states.

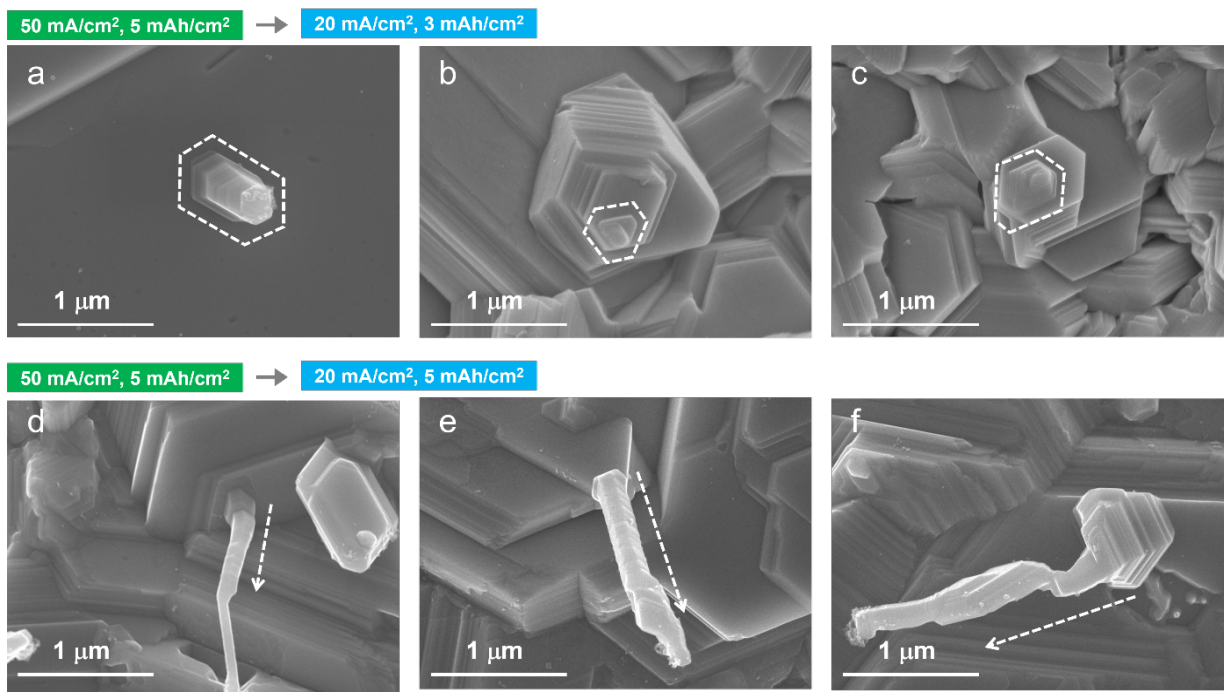

**Figure S9.** SEM images showing representative spiral features in the filamentous Zn deposits under $C_0 = 0.3$ M and 50 mA/cm² with a capacity of 5 mAh/cm², followed by an additional deposition at 20 mA/cm² with a capacity of (a-c) 3 and (d-f) 5 mAh/cm². It is clear that zinc deposits preferentially grow out of the basal plane when the current density decreases.

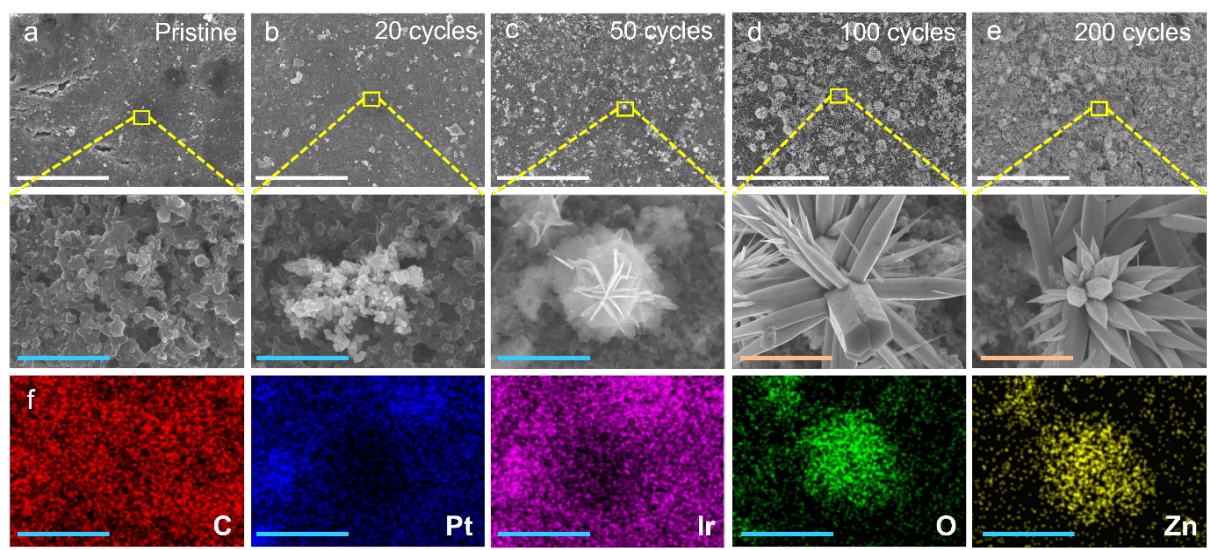

**Figure S10.** SEM images of cathodes for (a) pristine, and after the (b) 20th, (c) 50th, (d) 100th, and (e) 200th cycles under $C_0 = 0.3$ M and $i = 20$ mA/cm². (f) EDS mapping of the magnified structure in panel (c). The white, orange, and blue bars indicate 50, 2, and 1 μm, respectively.

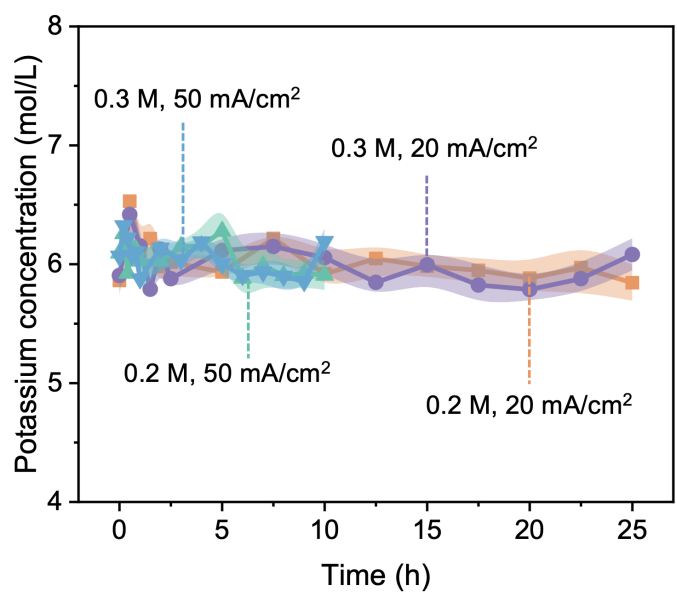

**Figure S11.** Potassium concentration remains almost constant at ~6 mol/L in electrolyte at the end of different charging process under different running conditions. Shaded areas indicate the standard deviations.

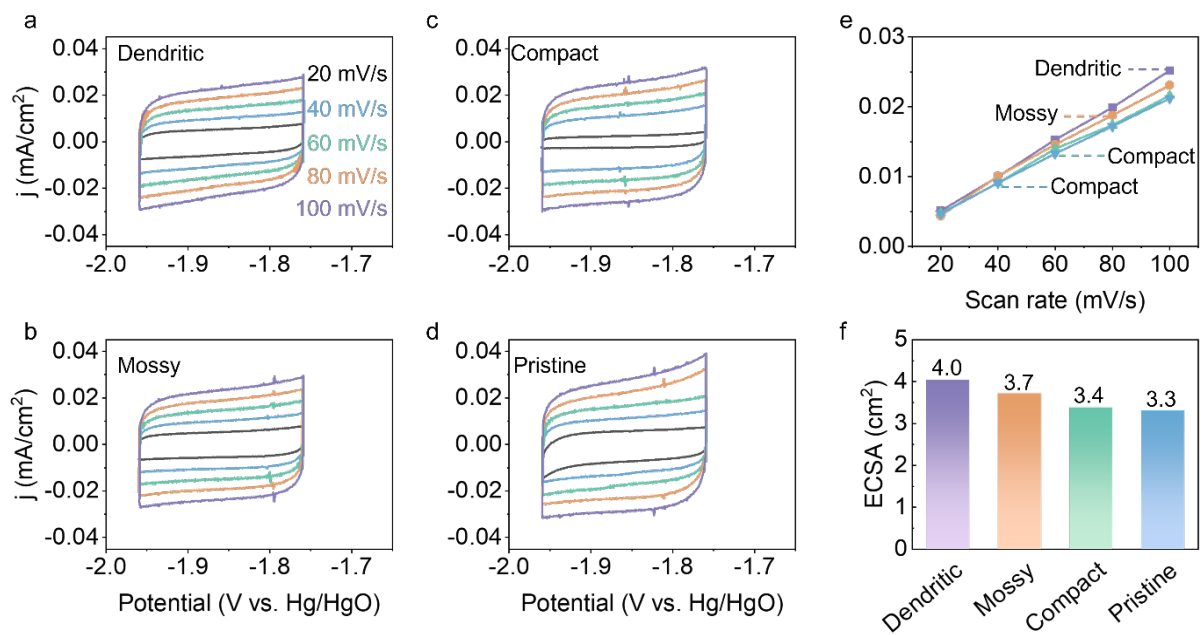

**Figure S12**. CV curves for (a) dendritic, (b) mossy, (c) compact, and (d) pristine zinc under different scan rates in 1 M KOH. (e) Current density versus scan rate at -1.85 V according to panels (a-d). Each slope represents the corresponding $C_{dl}$. (f) ECSA for different zinc anodes.

**References**

1. Blöchl, P. E., Projector Augmented-Wave Method. *Physical Review B* **1994**, *50*, 17953-17979.

2. Perdew, J. P.; Burke, K.; Ernzerhof, M., Generalized Gradient Approximation Made Simple. *Phys. Rev. Lett.* **1996**, *77*, 3865-3868.

3. Kresse, G.; Furthmüller, J., Efficient Iterative Schemes for Ab Initio Total-Energy Calculations Using a Plane-Wave Basis Set. *Physical Review B* **1996**, *54*, 11169-11186.

4. Grimme, S.; Antony, J.; Ehrlich, S.; Krieg, H., A Consistent and Accurate Ab Initio Parametrization of Density Functional Dispersion Correction (Dft-D) for the 94 Elements H-Pu. *J. Chem. Phys.* **2010**, *132*, 154104.

5. Henkelman, G.; Uberuaga, B. P.; Jónsson, H., A Climbing Image Nudged Elastic Band Method for Finding Saddle Points and Minimum Energy Paths. *J. Chem. Phys.* **2000**, *113*, 9901-9904.

6. Müller, S.; Holzer, F.; Haas, O., Optimized Zinc Electrode for the Rechargeable Zinc–Air Battery. *J. Appl. Electrochem.* **1998**, *28*, 895-898.

7. Kim, H.-W.; Lim, J.-M.; Lee, H.-J.; Eom, S.-W.; Hong, Y. T.; Lee, S.-Y., Artificially Engineered, Bicontinuous Anion-Conducting/-Repelling Polymeric Phases as a Selective Ion Transport Channel for Rechargeable Zinc–Air Battery Separator Membranes. *J. Mater. Chem. A* **2016**, *4*, 3711-3720.

8. Lee, D. U.; Park, H. W.; Park, M. G.; Ismayilov, V.; Chen, Z., Synergistic Bifunctional Catalyst Design Based on Perovskite Oxide Nanoparticles and Intertwined Carbon Nanotubes for Rechargeable Zinc–Air Battery Applications. *ACS Appl. Mater. Interfaces* **2015**, *7*, 902-910.

9. Meng, F.; Zhong, H.; Bao, D.; Yan, J.; Zhang, X., In Situ Coupling of Strung Co4n and Intertwined N–C Fibers toward Free-Standing Bifunctional Cathode for Robust, Efficient, and Flexible Zn–Air Batteries. *J. Am. Chem. Soc.* **2016**, *138*, 10226-10231.

10. Qian, Y.; Hu, Z.; Ge, X.; Yang, S.; Peng, Y.; Kang, Z.; Liu, Z.; Lee, J. Y.; Zhao, D., A Metal-Free Orr/Oer Bifunctional Electrocatalyst Derived from Metal-Organic Frameworks for Rechargeable Zn-Air Batteries. *Carbon* **2017**, *111*, 641-650.

11. You, T.-H.; Hu, C.-C., Designing Binary Ru–Sn Oxides with Optimized Performances for the Air Electrode of Rechargeable Zinc–Air Batteries. *ACS Appl. Mater. Interfaces* **2018**, *10*, 10064-10075.

12. Chen, B.; He, X.; Yin, F.; Wang, H.; Liu, D.-J.; Shi, R.; Chen, J.; Yin, H., Mo-Co@N-Doped Carbon (M = Zn or Co): Vital Roles of Inactive Zn and Highly Efficient Activity toward Oxygen Reduction/Evolution Reactions for Rechargeable Zn–Air Battery. *Adv. Funct. Mater.* **2017**, *27*, 1700795.

13. Wei, L.; Karahan, H. E.; Zhai, S.; Liu, H.; Chen, X.; Zhou, Z.; Lei, Y.; Liu, Z.; Chen, Y., Amorphous Bimetallic Oxide–Graphene Hybrids as Bifunctional Oxygen Electrocatalysts for Rechargeable Zn–Air Batteries. *Adv. Mater.* **2017**, *29*, 1701410.

14. Fu, G.; Wang, J.; Chen, Y.; Liu, Y.; Tang, Y.; Goodenough, J. B.; Lee, J.-M., Exploring Indium-Based Ternary Thiospinel as Conceivable High-Potential Air-Cathode for Rechargeable Zn–Air Batteries. *Adv. Energy Mater.* **2018**, *8*, 1802263.

15. Ma, H.; Wang, B.; Fan, Y.; Hong, W., Development and Characterization of an Electrically Rechargeable Zinc-Air Battery Stack. *Energies* **2014**, *7*, 6549-6557.

16. Zhang, Z.; Zhou, D.; Li, Z.; Zhou, L.; Huang, B., Preparation and Properties of a Zno/Pva/B-Cd Composite Electrode for Rechargeable Zinc Anodes. *ChemistrySelect* **2018**, *3*, 10677-10683.